\documentclass[11pt]{article}
\usepackage{amssymb}
\usepackage{epsfig}
\newlength{\bredde}
\def\slash#1{\settowidth{\bredde}{$#1$}\ifmmode\,\raisebox{.15ex}{/}
\hspace*{-\bredde} #1\else$\,\raisebox{.15ex}{/}\hspace*{-\bredde} #1$\fi}
\newcommand{\be}{\begin{equation}}
\newcommand{\ee}{\end{equation}}
\newcommand{\bea}{\begin{eqnarray}}
\newcommand{\eea}{\end{eqnarray}}
\newcommand{\nn}{\nonumber}

\newcommand{\al}{\alpha}

\newcommand{\ep}{\epsilon}
\newcommand{\sect}[1]{\setcounter{equation}{0}\section{#1}}

\def\zc{{z^\ast}}
\def\tr{{\mbox{tr}}}
\def\re{{\Re\mbox{e}}}
\def\im{{\Im\mbox{m}}}
\def\Jd{J^\dagger}

\begin{document}
\topmargin -1.4cm
\oddsidemargin -0.8cm
\evensidemargin -0.8cm
\title{\Large{
{\bf Microscopic universality of complex matrix model correlation functions 
at weak non-Hermiticity
}}}

\vspace{1.5cm}

\author{~\\{\sc G. Akemann}\\~\\
Service de Physique Th\'eorique, CEA/DSM/SPhT\\
Unit\'e de recherche associ\'ee au CNRS\\
CEA/Saclay\\
F-91191 Gif-sur-Yvette C\'edex, France
}
\date{}
\maketitle
\vfill
\begin{abstract}
The microscopic correlation functions of non-chiral random matrix models with 
complex eigenvalues are analyzed for a wide class of non-Gaussian measures. In 
the large-$N$ limit of weak non-Hermiticity, where $N$ is the size of the 
complex matrices, we can prove that all $k$-point 
correlation functions including an arbitrary number of Dirac mass terms are 
universal close to the origin. To this aim we establish the universality of 
the asymptotics of orthogonal polynomials in the complex plane. 
The universality of the correlation functions then follows from that of the 
kernel of orthogonal polynomials and a mapping of massive to massless 
correlators. 
\end{abstract}
%PACS 12.38.Lg, 11.15.Pg, 05.40.-a\\
%Keywords: complex random matrix models, universality, weak non-Hermiticity,
% QCD Dirac spectrum
\vfill

\begin{flushleft}
SPhT T02/066\\
\end{flushleft}
\thispagestyle{empty}
\newpage

\renewcommand{\thefootnote}{\arabic{footnote}}
\setcounter{footnote}{0}

%%%%%%%%%%%%%%%%%%%%%%%%%%%%%%%%%%%%%%%%%%%%%%%%%%%%%%%%%%%%%%%%%%%%%%%%%%%
\sect{Introduction}\label{intro}

The Universality of correlation functions is one of the crucial properties 
of random matrix models. In any of their applications,
where we refer to \cite{GMW} for a review, the mapping between 
the underlying field theory and the matrix model as an effective model is only 
dictated by the global symmetries of the field theory. 
On the matrix model side the invariance under this symmetry, 
given for example by unitary transformations $U(N)$, does not entirely 
fix the matrix model action 
and allows for a large variety of terms. The statement of universality is that
in general the quantities of interest such as 
matrix eigenvalue correlation functions 
are the same for a certain 
class of matrix model actions in the large-$N$ limit. 
The simplifying choice of for example a Gaussian therefore does not matter. 

One has to distinguish between different kinds of universality. In the 
macroscopic large-$N$ 
limit the eigenvalues of the random matrices which for example
correspond to energy levels are kept unscaled.
In this limit the oscillatory behavior of the correlation functions is 
smoothed and all two- and higher $k$-point functions are universal, to 
all orders in an expansion in $1/N^2$ \cite{Amb}. 
Such universality results find applications in Quantum Gravity \cite{Amb}, 
condensed matter physics of conductors \cite{Been} or more recently in the 
correspondence between supersymmetric Yang-Mills theory and low-energy 
String Theory \cite{ADIV}. 

In this work we concentrate on a different universality called microscopic. 
Here, eigenvalues and correlation functions are rescaled with the size of the 
matrices $N$, magnifying a particular part of the energy spectrum. 
Microscopic universality states that correlation functions are the same 
for a class of matrix model 
actions when measured with respect to the same mean level 
spacing between the eigenvalues. The eigenvalue spectrum 
of random matrix models is typically compact, similar for example to 
lattice gauge theories where large momenta are cut off due to 
the finite lattice spacing. 
In the simplest case the matrix model spectral density given 
by the Wigner semi-circle on an interval. 
We distinguish between the origin, bulk and edge region
of the spectrum where random matrix model universality was proven in 
\cite{ADMN}, \cite{BZ} and \cite{KF} respectively 
(for a review see \cite{KFrev}). The origin of the spectrum 
is particularly interesting for applications to chiral symmetry breaking 
in low-energy Quantum Chromodynamics 
(QCD) \cite{Jac} because of the relation between the order 
parameter, the chiral condensate, and the spectral density of QCD Dirac 
operator eigenvalues at the origin \cite{BC}. 
Here, universality has been shown for a 
large class of polynomial \cite{ADMN} and non-polynomial potentials 
\cite{AVII}, at criticality \cite{ADMNII} 
and in the presence of massive Dirac terms \cite{DN}. 
For classes with orthogonal and symplectic symmetry proofs including
massless and massive Dirac 
terms have been given in \cite{SV} and \cite{AK} respectively.

All we have 
said so far is valid for random matrices with real eigenvalues only. 
However, more recently applications for models with complex eigenvalues 
have occured as well. To give some examples we mention 
the fractional Quantum Hall Effect \cite{PdF}, 
two-dimensional charged plasmas \cite{Janco,FJ},
QCD with chemical potential \cite{Steph}, 
as well as the inverse potential problem in integrable systems 
\cite{Wieg}. 
We note that the chiral ensembles also contain complex matrices \cite{Jac,Amb} 
but remain in the class of real eigenvalue models.

The number of different symmetry classes among complex matrix models is 
particularly rich \cite{Denis}. Furthermore, 
one has to distinguish between the large-$N$ limit of strongly and weakly  
non-Hermitian matrices. While in the former case results date back to 
Ginibre \cite{Gin} the latter limit has been defined rather recently 
\cite{So} and correlation functions have been calculated in 
\cite{FKS,A01,EK,A02}.
The class of microscopic correlators in the weak non-Hermiticity limit 
plays an important r\^ole as it 
permits to interpolate between the correlations  of real eigenvalues 
and those of complex eigenvalues at strong non-Hermiticity. 
Our studies are devoted to the microscopic origin scaling limit at weak
non-Hermiticity, where up to date almost nothing was known about universality. 
The only exception is the microscopic spectral density which has been shown
to be universal in \cite{FKS98} using supersymmetric techniques 
for ensembles of sparse matrices with 
independently distributed entries and finite second moment.
While supersymmetric techniques become quite involved 
for higher correlators we will exploit the existence of orthogonal polynomials
in the complex plane \cite{PdF,A02} 
and analyze their asymptotic properties. Due to their universality and the 
power of the method of orthogonal polynomials \cite{Mehta} we can 
derive the universality of all $k$-point functions.
Along the way of our analysis we will encounter similarities between 
the complex matrix model and the two-matrix model with two independent matrices
as being studied for example in \cite{Bert}.

We will restrict ourselves to the simplest model, the complex extension 
of the Hermitian one-matrix model as introduced and solved in \cite{PdF,FKS}. 
In \cite{A01} the same model including an arbitrary number of 
Dirac masses was solved in the limit of weak and strong\footnote{In 
the massless limit the one- and two-point function 
were already calculated in \cite{Janco} at maximally strong non-Hermiticity, 
using the picture of a charged plasma. The arbitrary charge at the origin there
corresponds to massless Dirac terms in \cite{A01}.} non-Hermiticity 
and its relevance for three dimensional QCD with chemical potential was 
pointed out. 
In order to show also microscopic universality of the chiral 
model with complex eigenvalues 
introduced and solved very recently in \cite{A02}, 
we would need to know higher order $1/N$ corrections to our method.
The chiral 
model is relevant for QCD with chemical potential in four dimensions.
Apart from its universality it would be very interesting to see if it is indeed
in the same class of models as \cite{Steph}. 

The article is organized as follows. In Section \ref{model} we define our 
non-Gaussian 
complex matrix model together with all relevant quantities at finite-$N$. 
In the following Section \ref{OPchap} a second order differential 
equation for the 
orthogonal polynomials in the complex plane is derived in order to determine 
their asymptotic behavior. This leads to the universality of all $k$-point 
correlation functions of complex eigenvalues in the microscopic origin 
scaling limit at weak non-Hermiticity in Section \ref{Uni}. Our
conclusions and future prospects are presented in Section \ref{con}.

%%%%%%%%%%%%%%%%%%%%%%%%%%%%%%%%%%%%%%%%%%%%%%%%%%%%%%%%%%%%%%%%%%%%%%%%%%%

\sect{The model}\label{model}

The partition function of the matrix model we wish to study is defined as 
\bea
{\cal Z}_N^{(2N_f)} &\equiv& \int dJdJ^{\dagger}
\prod_{f=1}^{N_f}%\left|\det[J-im_f]\right|^2 
    \det[(J-im_f)(\Jd+im_f)] 
\ \exp\left[-N\tr\, V(J,\Jd) \right] \ , \label{Z}\\
 V(J,\Jd)&\equiv& \frac{1}{1-\tau^2}\left(J\Jd-\frac{\tau}{2}(J^2
  +J^{\dagger\, 2})\right)
\ +\ \frac12 \sum_{k=2}^d \frac{g_{2k}}{2k} (J^{2k}+ J^{\dagger\, 2k}) \ .
\label{V}
\eea
The complex $N\times N$ matrix $J$ is parameterized as 
\be
J\ \equiv\ H +  \sqrt{\frac{1-\tau}{1+\tau}}\ iA \ , \ \ \ \tau\in[0,1]\ , 
\label{J}
\ee
where $H$ and $A$ are Hermitian.
The potential $ V(J,\Jd)$ consists of a quadratic, Gaussian part 
depending on the parameter $\tau$ as well as of an arbitrary but fixed number 
of higher even powers in $J$ and $\Jd$ of maximal degree $2d$. 
The Gaussian part is fixed 
by requiring a Gaussian weight for both matrices $H$ and $A$ at equal variance
$(1+\tau)/(2N)$. When enlarging the measure to study universality we have to 
make a choice which terms we allow. Instead of allowing for arbitrary even 
potentials for the matrices $H$ and $A$ respectively, inserting $J$ and 
$\Jd$ according to eq. (\ref{J}), we have only introduced even powers of 
in $J$ and $\Jd$. The reason why we have to exclude higher order powers of 
mixed terms, $(J\Jd)^{k\geq2}$, goes as follows. In order to write eq. 
(\ref{Z}) in terms of eigenvalues we have to do the following unitary 
transformation, $J=U(Z+R)U^\dagger$ \cite{FKS}. Here $U$ is unitary, 
$Z=\mbox{diag}(z_1,\ldots,z_N)$ is the diagonal matrix of complex eigenvalues 
of $J$, and $R$ is a strictly upper triangular matrix\footnote{We could also 
diagonalize $J$ by a biunitary transformation, $J=U\Lambda V^\dagger$.
This would diagonalize products $(J\Jd)^k$ %as in \cite{AKM}, 
but not $J^2$ and $J^{\dagger\, 2}$ in the 
Gaussian part.}. Since products of strictly upper triangular and 
diagonal matrices remain traceless, the $R$-dependence drops out in all powers
$\tr J^{2k}$ and similarly in its Hermitian conjugate. In 
$\tr(J\Jd)=\tr(ZZ^\dagger + RR^\dagger)$ the $R$-dependence persists but 
can be integrated out because it is Gaussian. Adding higher powers 
$\tr(J\Jd)^{k}$ mixed terms between $Z$, $Z^\dagger$ 
and $RR^\dagger$ would occur and the 
matrix $R$ could no longer be integrated out. This is why we restrict 
ourselves to the form of the potential eq. (\ref{V}). A similar class of 
potentials has also been considered in \cite{Wieg}.
In terms of complex eigenvalues $z_{j=1,\ldots,N}$ 
the partition function eq. (\ref{Z}) thus reads
\bea
{\cal Z}_N^{(2N_f)} &=& 
\int \prod_{j=1}^N
\left(d^2\!z_j %dz_j^\ast
\ \prod_{f=1}^{N_f}|z_j-im_f|^2%(z^\ast+im_f)
\  \mbox{e}^{-N V(z_j,z_j^\ast)}\right)
\left|\Delta_N(z_1,\ldots,z_N)\right|^2, 
\label{Zev}\\
 V(z,z^\ast)&=& \frac{1}{1-\tau^2}\left(zz^\ast-\frac{\tau}{2}(z^2
  +z^{\ast\,2})\right)
\ +\ \frac12\sum_{k=2}^d \frac{g_{2k}}{2k} (z^{2k}+ z^{\ast\, 2k}) \ ,
\label{Vev}
\eea
with 
$\Delta_N(z_1,\ldots,z_N)=\prod_{k>l}^N (z_k-z_l)$ being 
the Vandermonde determinant. We have dropped the constants coming from the 
integration over the matrices $U$ and $R$.
The $k$-point correlation functions among 
the $N$ eigenvalues $z_j$ are defined in the usual way \cite{Mehta}:
\bea
R_N^{(2N_f)}(z_1,\ldots,z_k) &\equiv& 
 \frac{N!}{(N-k)!}\frac{1}{{\cal Z}_N^{(2N_f)}} \label{Rkdef} \\
&\times&\int d^2\!z_{k+1}\ldots d^2\!z_N \prod_{j=1}^N\left(
\prod_{f=1}^{N_f} |z_j-im_f|^2\ \mbox{e}^{-NV(z_j,z_j^\ast)}\right) 
|\Delta_N(z_1,\ldots,z_N)|^2\ . \nn
\eea
As it has already been pointed out in \cite{AK,A01} the correlation functions 
with and without the $N_f$ quark masses are related. Absorbing the mass terms 
into the Vandermonde the following relation\footnote{At finite-$N$ the 
relation holds exactly when the $N$-dependence in the weight is absorbed 
into the eigenvalues, e$^{-V(z,\zc)}$.} 
has been derived \cite{A01}\footnote{In eq. (2.21) of ref. \cite{A01} 
an additional $N$ is missing in the subscript of the denominator.}:
\be
R_N^{(2N_f)}(z_1,\ldots,z_k) \ =\ 
\frac{R_{N+N_f}^{(0)}(z_1,\ldots,z_k,im_1,\ldots,im_{N_f})}
{R_{N+N_f}^{(0)}(im_1,\ldots,im_{N_f})} \ .
\label{Rkmass}
\ee
Consequently we only have to calculate the correlators at $N_f=0$ and prove 
their universality. The universality of the massive correlation functions then
follows from eq. (\ref{Rkmass}) at large-$N$.

The $k$-point correlation functions eq. (\ref{Rkdef}), massless or massive, 
can be obtained in the standard way \cite{Mehta} using orthogonal polynomials. 
These are defined in the complex plane as 
\be
\int d^2\!z\ \mbox{e}^{-N V(z,\zc)}
P_k(z)P_l(\zc) \ =\ \delta_{kl} \ ,
\label{OPdef}
\ee
were from now on we suppress the superscript $(2N_f=0)$. The kernel of 
orthogonal polynomials is then given by 
\be
K_N(z_1,z_2^\ast) \ \equiv\ 
\mbox{e}^{-\frac{N}{2}(V(z_1,z_1^\ast)+V(z_2,z_2^\ast))}
\sum_{l=0}^{N-1} P_l(z_1)P_l(z_2^\ast)\ ,
\label{kerneldef}
\ee
and the correlation functions follow from it \cite{Mehta}
\be
R_N(z_1,\ldots,z_k) \ =\
\det_{i,j=1,\ldots,k}\left[K_N(z_i,z_j^\ast)\right]\ .
\label{Rker}
\ee
Already at finite-$N$ the orthogonal polynomials eq. (\ref{OPdef}) possess 
a matrix or eigenvalue representation for an arbitrary potential $V(J,\Jd)$ 
eq. (\ref{V}). 
For that purpose we introduce a parameter dependent  partition function 
with matrices of smaller size $n\times n$, where $n\leq N$:
\be
{\cal Z}_n(t) \ \equiv\ \int (dJdJ^{\dagger})_{n\times n}
\ \exp\left[-\frac{n}{t}\tr V(J,\Jd) \right]  \ .
\label{Zt}
\ee
As one can easily convince oneself the following representation for
orthogonal polynomials \cite{Sze} also holds in the complex plane:
\be
\tilde{P}_n(z) \ =\   \frac{1}{{\cal Z}_n(t=\frac{n}{N})}
\int (dJdJ^{\dagger})_{n\times n} \det(z-J)
\ \exp\left[-N\tr\, V(J,\Jd) \right] 
\ =\ \left\langle\,\det(z-J)\,\right\rangle_{{\cal Z}_n(t=\frac{n}{N})}\  .
\label{OPrepun}
\ee
Here, they are normalized as $\tilde{P}_n(z)=z^n+\ldots$ . In order to obtain 
orthonormal polynomials as in the definition 
(\ref{OPdef}) we divide by their norm, $h_n$,
to obtain
\be
P_n(z)\ =\ h_n^{-\frac12} \tilde{P}_n(z)\ =\ h_n^{-\frac12}
\left\langle\,\det(z-J)\,\right\rangle_{{\cal Z}_n(\frac{n}{N})}\  .
\label{OPrep}
\ee
For the issue of universality we will also have to keep track of the 
$(t=\frac{n}{N})$-dependence of the polynomials in the 
asymptotic large-$n$ limit keeping $t=\frac{n}{N}$ finite. 
We repeat that eqs. (\ref{OPdef}) -- (\ref{OPrep}) 
also hold in the presence of $2N_f$ masses in the measure.

The resolvent for the partition function eq. (\ref{Zt}) is defined as
\be
G_n(z;t)\ \equiv\  \frac{1}{{\cal Z}_n(t)}
 \int (dJdJ^{\dagger})_{n\times n}\ \frac{1}{n} \tr\frac{1}{z-J}
\ \exp\left[-\frac{n}{t}\tr V(J,\Jd) \right]  
\ = \ \left\langle\,\frac{1}{n} \tr\frac{1}{z-J}
\,\right\rangle_{{\cal Z}_n(t)} .
\label{Gt}
\ee
It can also be expressed through the 1-point function, the spectral density, by
\be
G_n(z;t)\ = \ \frac{1}{n}\int d^2\!w \,\frac{R_n(w)}{z-w}\ .
\label{GRn}
\ee 
This representation is an alternative to calculate in particular 
the macroscopic large-$n$ spectral density from a saddle point analysis. 
We can deduce the following relation between density and resolvent in the 
complex plane which is different for real eigenvalues (see eq. (\ref{GrhoR}) 
below):
\be
\frac{1}{\pi} 
\partial_{z^\ast}G_n(z;t)\ =\ \frac{1}{n}R_n(z)\ .
\label{GrhoC}
\ee

%%%%%%%%%%%%%%%%%%%%%%%%%%%%%%%%%%%%%%%%%%%%%%%%%%%%%%%%%%%%%%%%%%%%%%%%%%%%

\sect{The asymptotic of orthogonal polynomials
}\label{OPchap} 

In this section we will derive a differential equation for the orthogonal 
polynomials from the representation eq. (\ref{OPrep}). The universality 
of the asymptotic polynomials and in consequence of the correlation functions 
then follow from it. The fact that we will take the large-$n$ limit at weak 
non-Hermiticity will allow us to borrow some results from the proof in the 
real case \cite{ADMN} despite the fact that the microscopic correlations 
spread out into the complex plane in our case. The polynomials defined in 
eq. (\ref{OPdef}) have parity symmetry because of the even potential chosen. 
We will therefore aim at a second order differential equation that closes on 
polynomials of equal parity. 

We start by taking the derivative of $P_n(z)$ from eq.(\ref{OPrep}) 
\bea
\partial_z P_n(z) &=&  n h_n^{-\frac12}
 \left\langle \frac{1}{n}\tr\frac{1}{z-J}\ \det(z-J)
\right\rangle_{{\cal Z}_n(\frac{n}{N})} 
\label{P'}\\
&=& n G_n\left(z;t=\frac{n}{N}\right) P_n(z) \ +\ 
 nh_n^{-\frac12}\left\langle \frac{1}{n}\tr\frac{1}{z-J}\ \det(z-J)
\right\rangle^{con\!n}_{{\cal Z}_n(\frac{n}{N})} \ ,\nn
\eea
where we have introduced the connected part ($con\!n$) 
of the expectation value. 
At large-$n$ expectation values are known to factorize and the second 
term will be suppressed by $\frac{1}{n^2}$ with respect to the first. We could
in principle also determine the asymptotic of $P_n(z)$ from the above equation 
as it has been previously done in \cite{BZ} in the real case. However, in 
order to obtain the correct behavior, sine and cosine as it will turn out, 
we would have to know also subleading corrections in $N$
%of order ${\cal O}(\frac{1}{n})$ 
to fix their parity as in the real case \cite{BZ}. 
Furthermore it is known that for orthogonal polynomials 
of real eigenvalues an exact, finite-$n$ 
differential equation exists \cite{KFrev}
which only closes on the polynomials as an equation of second order. 
Taking one more derivative we arrive at
\bea
\partial^2_z P_n(z) &=&  
n \partial_zG_n\left(z;\frac{n}{N}\right) P_n(z) 
\ +\ n^2G_n\left(z;\frac{n}{N}\right)^2 P_n(z) \label{P''n}\\
&+&  nG_n\left(z;\frac{n}{N}\right)
 h_n^{-\frac12}\left\langle \frac{1}{n}\tr\frac{1}{z-J}\det(z-J)
\right\rangle^{con\!n}_{{\cal Z}_n(\frac{n}{N})} 
+ nh_n^{-\frac12}\partial_z
 \left\langle \frac{1}{n}\tr\frac{1}{z-J}\det(z-J)
\right\rangle^{con\!n}_{{\cal Z}_n(\frac{n}{N})}
\label{preP''}
\nn
\eea
which is exact at finite $n$. At large-$n$ the connected expectation
values in the second line are suppressed by $\frac{1}{n^2}$ as follows from 
general counting arguments in matrix models with unitary invariance. 
Nevertheless the last term is of ${\cal O}(1)$ since the 
derivative may produce another factor of $n$. We thus end 
up with the following second order differential equation:
\be
\partial^2_z P_n(z) \ =\   
\left[ 
n^2G_n\left(z;\frac{n}{N}\right)^2 
\ +\ 
n \partial_z G_n\left(z;\frac{n}{N}\right) 
\ +\ {\cal O}(1)
\right]P_n(z) 
\ .
\label{P''}
\ee
We now proceed exactly as in the universality proofs for real eigenvalues. 
The coefficient functions in front of the polynomials in eq. 
(\ref{P''}) are smoothed and thus replaced by their values in the 
macroscopic large-$n$ limit. The precise meaning of smoothing is defined as 
follows. The orthogonal polynomials $P_n(z)$ obey a so-called 
recursion relation where $zP_{n-1}(z)$ is reexpressed in terms of linear 
combinations of polynomials $P_{l}(z)$ for $l\leq n$. 
On the real line a three step recursion relation holds for arbitrary polynomial
potentials, which is no longer true in the complex plane. Here, the degree 
of the recursion explicitly depends on the degree $2d$ of the potential 
eq. (\ref{Vev}), similarly to the two-matrix model \cite{Bert}.  
Smoothing is defined by the assumption that the recursion coefficients $r_l$ 
approach a single function in the variable 
$t=\frac{l}{N}$. Let us give an example. For the Gaussian potential given by  
$V(z,z^\ast)=\frac{1}{1-\tau^2}
\left(zz^\ast-\frac{\tau}{2}(z^2+z^{\ast\,2})\right)$ the orthogonal 
polynomials are known to be Hermite polynomials in the complex plane 
\cite{PdF}, 
$P_n(z)=\sqrt{\frac{N}{\pi n!}}(1-\tau^2)^{-\frac14}\tau^{\frac{n}{2}} 
H\!e_n(z\sqrt{N/\tau})$,  
and they obey the following recursion relation\footnote{From eq. 
(\ref{recG}) one can
see that the Christoffel-Darboux formula for the 
kernel eq. (\ref{kerneldef}) no longer holds for $\tau<1$.} 
\be
z P^{Gaus\!s}_{n-1}(z) \ =\ \sqrt{\frac{n}{N}}\,P^{Gaus\!s}_n(z)
                      + \sqrt{\frac{n-1}{N}}\ \tau P^{Gaus\!s}_{n-2}(z)\ .
\label{recG}
\ee
Obviously the recursion coefficients $r_n=n/N$ approach a smooth limit, 
$r_{n,n\pm1,\pm2,...}\to r(t)=t$. 
Taking this limit we obtain the following 
expression for the Gaussian resolvent outside the support $\sigma$ of the 
complex eigenvalues
\be
G^{Gaus\!s}(z;t) \ =\ \frac{1}{t}\left( \frac{z}{2\tau} -
\sqrt{\frac{z^2}{4\tau^2}-\frac{1}{\tau}}\right) \ ,\ \ 
z\in\mathbb{C}\backslash \sigma\ ,
\label{GG}
\ee
which coincides with \cite{So} at $t=1$. Here, we have used the 
fact that the resolvent eq. (\ref{GRn}) can be expressed through the 
kernel via eq. (\ref{Rker}). Expanding the pole and 
smoothing the non-vanishing even powers, 
$\int d^2\!z \mbox{e}^{-nV/t}z^{2k}P_l(z)P_l(\zc)=(\tau t)^k{{2k}\choose{k}}$, 
where we can follow the calculation for real eigenvalues given 
e.g. in \cite{KFrev}, 
we arrive at eq. (\ref{GG}). It is the macroscopic resolvent outside the 
support as can be seen from the vanishing of eq. (\ref{GrhoC}),
$\partial_{z^\ast}G^{Gaus\!s}(z;t)=0$.
For our potential eq. (\ref{Vev}) the recursion relation (\ref{recG}) extends
down to $P_{n-2d+1}(z)$. 
We assume that the recursion coefficients have a smooth 
limit and that the procedure leads to the macroscopic resolvent, $G(z,t)$, 
and its derivative $\partial_z G(z,t)$ in eq. (\ref{P''}).
Because of the complicated form of the recursion relation for our general 
potential we use another derivation of the macroscopic resolvent using loop 
equations \cite{Amb}. 
Exploiting the invariance of the partition function eq. (\ref{Zt}) 
under the change of variables $J\to J +\ep/(z-J)$ and its Hermitian conjugate
($H.c.$) leads to the following equation, after requiring 
$\left.\partial_\ep{\cal Z}(t)\right|_{\ep=0}\!=0$:
\be
0\ =\ \left\langle \left( \tr \frac{1}{z-J}\right)^2\right\rangle_{{\cal Z}(t)}
-\ \left\langle\frac{n}{t} \tr \left[ (\partial_J V(J,J^\dagger)) \frac{1}{z-J}
\right]\right\rangle_{{\cal Z}(t)} +\ H.c.
\label{loop1} \ \ .
\ee
It contains mixed terms as $J^{\dagger\, 2k-1}/(z-J)$ which cannot be written 
solely in terms of eigenvalues, as explained in Section \ref{model}. 
This mixing can be partially removed for the Gaussian part of the potential
with the help of the identity 
\be
0\ =\ \left\langle 
\frac{n}{t} \tr \left[ (\partial_J V(J,J^\dagger)) \frac{1}{z^\ast-J^\dagger}
\right]\right\rangle_{{\cal Z}(t)}
+\ H.c.
\label{loop2} \ \ .
\ee
It follows from the invariance of ${\cal Z}(t)$ under  
$J\to J +\ep/(z^\ast-J^\dagger)$. Adding $\tau$ times eq. (\ref{loop1})
to eq. (\ref{loop2}) leads to 
\be
0\ =\ 
\tau\left\langle \left(\tr\frac{1}{z-J}\right)^2\right\rangle_{{\cal Z}(t)}
-\left\langle\frac{n}{t} \tr 
\left[ \mbox{\LARGE(}
J + \frac{1}{2}\sum_{k=2}^d g_{2k}(\tau J^{2k-1}+ J^{\dagger\, 2k-1})
\mbox{\LARGE)} 
\frac{1}{z-J}\right]
\right\rangle_{{\cal Z}(t)} +\ H.c.
\label{loop} \ ,
\ee
from which we easily recover eq. (\ref{GG}) at large-$n$ when restricting us
to the Gaussian. We cannot 
deduce the macroscopic resolvent $G(z,t)$ from eq. (\ref{loop}) 
for a generic potential as it still depends on the matrix $R$ from the 
diagonalization of $J$. To disentangle $J$ and $\Jd$ 
we would need more identities of the type eq. (\ref{loop2}). 
However, eq. (\ref{loop}) will be sufficient to obtain the macroscopic 
resolvent in the weak non-Hermiticity limit. It is defined by taking
$\tau\to 1$, as in the Hermitian limit, but with keeping the following 
product fixed:
\be
N(1-\tau^2) \equiv \al^2 \ .
\label{weaklim}
\ee
Whereas the appropriately rescaled microscopic quantities (see definition 
(\ref{rhomicrodef})) will depend on complex scaling variables and on
the parameter $\al$, the macroscopic quantities are projected 
onto the real axis because of $\tau\to1$. The smoothed resolvent $G(z;t)$ 
or the macroscopic spectral density, 
$\rho(z)\equiv\lim_{N\to\infty}N^{-1}R_N(z)$, 
only depend on $x=\re z$ since macroscopically $\im z$ vanishes.
This can be explicitly seen 
from eq. (\ref{GG}) where no $\al$ dependence occurs.
Furthermore,  from the defining relation between 
the microscopic and macroscopic density at the origin we have 
\be
\lim_{|\xi|\to\infty}\rho_S(\xi) \ =\ \rho(0)\ .
\label{mima}
\ee
Taking the results for a Gaussian potential from \cite{A01} which 
depend on $\xi,\al$ and the masses, it can be explicitly seen 
that the macroscopic density $\rho(0)$ at weak non-Hermiticity is only 
non-vanishing when taking the limit $|\xi|\to\infty$ along the real line
\footnote{Eq. (\ref{mima}) also holds at strong non-Hermiticity. There 
$\rho(0)$ is non-vanishing when taking the limit $|\xi|\to\infty$
along all directions, 
indicating that the support extends into the complex plane.}.
Therefore the support of the macroscopic density collapses from a two 
dimensional set to an
interval, e.g. $\sigma=[-2,2]$ in the Gaussian case (see eq. (\ref{GG})).
For the macroscopic quantities on the real line the relation (\ref{GrhoC}) 
is no longer valid. From the loop equation (\ref{loop}) at $\tau=1$ and
large-$n$ we have\footnote{Since the resolvent is no longer defined 
on the support $\sigma$ of real eigenvalues
we add a small imaginary part $\pm i\ep$.}
\bea
G(x\pm i\ep,t)
&=& \frac{1}{2t}V'(x) \ \mp\ i\pi\rho_t(x)\ ,\ \ x\in\sigma
\subset\mathbb{R} \ ,
\label{GrhoR}\\
\mbox{where}\ \ \ \ \ \ \ \ \ \ \ \ \ \ \ \ \ \ \ \ \ && \nn\\
V(x)&\equiv&\sum_{k=1}^d 
\frac{g_{2k}}{2k} x^{2k} \ ,\ \ g_2=1\ ,\label{Vx}\\
\rho_t(x) &\equiv&
\frac{1}{2\pi} \sum_{j=1}^m \frac{g_{2j}}{t} \sum_{k=0}^{j-1} 
{{2k}\choose{k}}
r(t)^{k} x^{2(j-k-1)}
\sqrt{4r(t)-x^2}  \, .
\label{Vrho}
\eea
All quantities are now real valued functions depending on $x=\re z$ 
only, with the above eqs. %(\ref{GrhoR}) and (\ref{Vrho}) 
being familiar from the Hermitian one-matrix model 
(e.g. in \cite{ADMN}). The quantity 
$r(t)$ is the limiting function of the recursion coefficients $r_n$ at 
large-$n$ in the three step recursion relation generalizing eq. (\ref{recG}). 
It is related to the endpoint $c$ of the eigenvalue support, $\sigma=[-c,c]$, 
through $c^2=4r(t)$. The determining equation following from the normalization 
condition $\int_\sigma dx \rho_t(x)=1$ is reading
\be
\frac12 \sum_{k=1}^m \frac{g_{2k}}{t} {{2k}\choose{k}}r(t)^{k}\ =\ 1\ .
\ee
With all ingredients supplied we can finally take 
the microscopic scaling limit of the asymptotic differential equation 
(\ref{P''}) at the origin. It is defined as 
\be
Nz\ =\ N(\re z +i\im z) \ \equiv\  \xi \ \ .
\label{microlim}
\ee
In this limit the complex eigenvalues $z$ approach zero while $\xi$ is 
kept fixed. The rescaling of the microscopic correlators is given by 
\be
\rho_S(\xi_1,\ldots,\xi_k) \ \equiv\ \lim_{N\to\infty} N^{-2k} 
R_N(N^{-1}\xi_1,\ldots,N^{-1}\xi_k) \ .
\label{rhomicrodef}
\ee
We take the smoothed  eq. (\ref{P''}) only depending on the macroscopic 
resolvent $G(z,t)$. The latter is projected to the real axis, 
due to the weak non-Hermiticity limit eq. (\ref{weaklim}). At the same time 
we perform the microscopic limit (\ref{microlim}),  rescaling the 
differential equation by $N^{-2}$ to obtain 
\be
\partial^2_\xi P_t(\xi) \ =\   
t^2G\left(0\pm i\ep;t\right)^2 P_t(\xi) \ .
\label{P''mic}
\ee
Here we have already removed the $\ep$ from the argument of the polynomials 
since they are analytic. The subscript $t$ indicates that they 
explicitly depend on $t=\frac{n}{N}\in(0,1]$ which is kept finite at
large-$N$. Taking the sum of eq. (\ref{P''mic}) for both signs and using 
eq. (\ref{GrhoR}) we finally arrive at
\be
\partial^2_\xi P_t(\xi) \ =\   
-\ t^2\pi^2 \rho_t(0)^2P_t(\xi) \ , 
\label{P''rho}
\ee
where the contribution from the potential $V'(0)$ vanishes.
The following form for the asymptotic polynomials thus holds:
\bea
\lim_{n,N\to\infty;\ \tau\to1} 
P_n\left(z=\frac{\xi}{N}\right)\ \to \ P_t(\xi)\ =\ 
\left\{
\begin{array}{ll}
f(t)\cos(t\pi\rho_t(0)\xi) & n\ \mbox{even}\\
g(t)\,\sin(t\pi\rho_t(0)\xi) & n\ \mbox{odd}\\
\end{array}
\right.\ \ ,\ t=\frac{n}{N}\ ,
\label{Pasymp}
\eea
where we have used their parity to select the 
appropriate solution. The normalization constants $f(t)$ and $g(t)$ 
still have to be determined. Our universal parameter $\rho_t(0)$
is the $t$-dependent macroscopic spectral density for the partition 
function eq. (\ref{Zt}) with potential $V(x)/t$ from eq. (\ref{Vx}). 
At $t=1$ it coincides with the macroscopic density $\rho(0)$ of
our model eq. (\ref{Z}) in the Hermitian limit (times a delta-function in 
$\im z$). 
We now introduce the following abbreviation,
\be
\pi t\rho_t(0) \ \equiv\ u(t) \ ,
\label{urho}
\ee
identifying the function $u(t)=\int_0^tds/(2\sqrt{r(s)})$ as it occurs in the 
universality proof of ref. \cite{ADMN}.  
The asymptotics of orthogonal polynomials in eq. (\ref{Pasymp}) is thus 
the same as that of the polynomials on the real line \cite{ADMN}, 
replacing the argument by a complex scaling variable $\xi$. 
However, the unknown coefficients $f(t)$ 
and $g(t)$ may still depend on the potential $V(x)$ in a complicated 
fashion. In particular, they also depend on the weak non-Hermiticity parameter 
$\al$ which is not present for real eigenvalues. Another difference is that 
because of the lack of a Christoffel-Darboux formula the correlation functions 
from eq. (\ref{Rker}) do not only depend on $P_t(\xi)$ at $t=1$ but on an 
integral over $t$ \cite{FKS,A01}. It is therefore a non-trivial task to show 
that the correlation functions of a Gaussian and our potential eq. (\ref{Vev}) 
agree.

%%%%%%%%%%%%%%%%%%%%%%%%%%%%%%%%%%%%%%%%%%%%%%%%%%%%%%%%%%%%%%%%%%%%%%%%%%%%

\sect{Universality of all correlation functions
}\label{Uni}

As a first step we calculate the prefactors $f(t)$ and $g(t)$ of the 
asymptotic polynomials eq. 
(\ref{Pasymp}) by taking the microscopic limit (\ref{microlim}) of the 
orthogonality relation (\ref{OPdef}). By keeping $t=\frac{k}{N}$ and 
$t'=\frac{l}{N}$ fixed at large-$N$ the Kronecker 
$\delta_{kl}$ becomes a delta-function.
On the left hand side we have to introduce microscopic variables replacing
$z=\frac{\xi}{N}$. In the potential eq. (\ref{Vev}) 
terms of higher order than quadratic are suppressed 
in the limits  (\ref{weaklim}) and (\ref{microlim}), 
leading to $\exp[-\frac{2}{\al^2}(\im \xi)^2]$. 
We arrive at 
\be
\int d\re\xi\,d\im\xi\ 
\mbox{e}^{-\frac{2}{\al^2}(\Im\mbox{\scriptsize m} \xi)^2}
P_t(\xi)P_{t'}(\xi^\ast) \ =\ \frac1N\delta(t-t') \ .
\label{OPNrel}
\ee
Inserting the asymptotic form of the polynomials (\ref{Pasymp}) we first 
perform the integral over the real part,
\be
\int_{-\infty}^\infty d\re \xi \left\{
\begin{array}{c}
\cos(u(t)\xi)\cos(u(t')\xi^\ast) \\
\sin(u(t)\xi)\sin(u(t')\xi^\ast) \\
\end{array}
\right\} \ =\ \mbox{\LARGE[}
\delta(u(t)-u(t'))\ +\ \delta(u(t)+u(t'))\mbox{\LARGE]} \pi 
\cosh(2u(t)\im \xi) \ .
\ee
The second term can be dropped because of the positivity of the spectral 
density and thus of $u(t)$. The equation for the prefactors 
then reads
\be
\frac{1}{N^2}
\frac{\pi}{u(t)'}\delta(t-t') 
\left\{
\begin{array}{c}
|f(t)|^2\\
|g(t)|^2\\
\end{array}
\right\}
\int_{-\infty}^\infty d\im \xi\, 
\mbox{e}^{-\frac{2}{\al^2}(\Im\mbox{\scriptsize m} \xi)^2}
\cosh(2u(t)\im \xi)
\ =\ \frac1N \delta(t-t') ,
\ee
where the derivative $u(t)'$ occurs. It leads to the following result
\be
f(t)\ =\ g(t)\ =\ \left[\sqrt{2}\,Nu(t)'\al^{-1}\pi^{-\frac32}
\right]^{\frac12} \mbox{e}^{-\frac{\al^2}{4}u(t)^2} \ .
\ee
The constant phase of the two functions which is arbitrary 
has been set to unity. The final result 
for the asymptotics of the polynomials is reading
\bea
\lim_{n,N\to\infty;\ \tau\to1} 
P_n\left(z=\frac{\xi}{N}\right)%\ \to \ P_t(\xi)\ =\ 
=\left[\sqrt{2}\,Nu(t)'\al^{-1}\pi^{-\frac32}
\right]^{\frac12} \mbox{e}^{-\frac{\al^2}{4}u(t)^2}
\left\{
\begin{array}{ll}
\cos(u(t)\xi) & n\ \mbox{even}\\
\sin(u(t)\xi) & n\ \mbox{odd}\\
\end{array}
\right. ,\ t=\frac{n}{N}\ .
\label{OPfinal}
\eea
The evaluation of the microscopic kernel 
defined as 
$K_S(\xi_1,\xi^\ast_2)\equiv \lim_{N\to\infty}N^{-2}K_N(\xi_1/N,\xi^\ast_2/N)$
is now straight forward:
\bea
K_S(\xi_1,\xi^\ast_2) 
&=& 
 \mbox{e}^{-\frac{1}{\al^2}((\Im\mbox{\scriptsize m} \xi_1)^2
+(\Im\mbox{\scriptsize m} \xi^\ast_2)^2)}
\frac{2}{\al\pi}\int_0^1 \frac{dt}{\sqrt{2\pi}}\ u(t)'
\mbox{e}^{-\frac{\al^2}{2}u(t)^2}
[\, \cos(u(t)\xi_1) \cos(u(t)\xi^\ast_2) \nn\\
&&\ \ \ \ \ \ \ \ \ \ \ \ \ \ \ \ \ \ \ \ \ \ 
\ \ \ \ \ \ \ \ \ \ \ \ \ \ \ \ \ \ \ \ \ \ 
\ \ \ \ \ \ \ \ \ \ \ \ \ \ \ \ \ \ \ + 
\ \sin(u(t)\xi_1)\sin(u(t)\xi^\ast_2)\,]
\nn\\
&=&  \mbox{e}^{-\frac{1}{\al^2}((\Im\mbox{\scriptsize m} \xi_1)^2
+(\Im\mbox{\scriptsize m} \xi^\ast_2)^2)}
\frac{2}{\al\pi}\int_0^{\pi\rho(0)} \frac{du}{\sqrt{2\pi}}\ 
\mbox{e}^{-\frac{\al^2}{2}u^2}
\cos\left(u(\xi_1-\xi_2^\ast)\right) \ .
\label{kerfin}
\eea
Here, we have replaced the sum in eq. (\ref{kerneldef}) by an integral, 
$\sum_{l=0}^{N-1}=N\int_0^1dt$ and we have substituted $t\to u(t)$. From eq. 
(\ref{urho}) $u(0)$ vanishes at the boundary.
Our result eq. (\ref{kerfin}) exactly coincides with the microscopic kernel 
calculated in \cite{FKS} for a Gaussian potential and is thus universal.
The known universal sine-kernel \cite{BZ} is reproduced in the limit $\al\to0$.
The universal parameter is the macroscopic density of the Hermitian model 
$\rho(0)$
at the origin and it resumes the dependence on all the coupling constants 
in our potential eq. (\ref{Vev}) via eq. (\ref{Vrho}) at $t=1$. 
The derivation presented here is an alternative to the one given  
in \cite{FKS98} for a Gaussian only, 
where an integral representation of the Hermite polynomials 
was used. A generalization of such a representation is not easy to obtain 
for our general potential considered.

The universality of the microscopic correlation functions at weak 
non-Hermiticity follows from eq. (\ref{kerfin}) 
by simply inserting eq. (\ref{Rker})
into the definition 
(\ref{rhomicrodef}). In taking the microscopic large-$N$ limit of eq. 
(\ref{Rkmass}) it trivially carries over to the massive correlation 
functions as well, 
as it has already been pointed out in \cite{A01}. For explicit 
expressions of the massive correlation functions we also refer to \cite{A01}.

In \cite{FKS} the microscopic correlation functions at weak non-Hermiticity 
away from the origin have also been calculated for a Gaussian. 
It is not easy to translate 
our analysis away from the origin. Already in the Gaussian case the leading 
order coefficient of the asymptotic expansion of the Hermite polynomials is 
not sufficient to reproduce \cite{FKS}. Since already in the Gaussian case 
we need the full finite-$n$ differential equation for the asymptotics of 
the Hermite polynomials 
the question of universality in the weak non-Hermiticity limit away from 
the origin is beyond the scope of this article.

Let us add two more remarks. As we have mentioned in Section \ref{OPchap} the 
whole formalism of orthogonal polynomials also holds in the presence of 
mass terms in the weight function. A natural question is why we 
could not directly derive the orthogonal polynomials including masses from 
the very same differential equation (\ref{preP''}). The reason is that the 
mass terms in eq. (\ref{Z}) are given as a product of determinants 
which at large-$N$ would also factorize and thus completely drop out of 
the representation eq. (\ref{OPrepun}). However, we know from the real 
case \cite{DN} that the massive and massless orthogonal polynomials 
are different. Therefore our leading order analysis in eq. (\ref{P''})
does not suffice in that case and we have circumvented this problem by  
the use of relation (\ref{Rkmass}). 
Our second remark concerns the universality of the chiral complex matrix model 
containing Laguerre polynomials in the complex plane, as very recently 
introduced in \cite{A02}. The modification inside the partition function 
eq. (\ref{Z}) to obtain these orthogonal 
polynomials can be simply achieved by adding a term $-(2a+1)\ln|z|/N$ to the 
potential in eq. (\ref{OPdef}) (and by squaring the arguments of the 
Vandermonde). However, going through the derivation we find 
that in eq. (\ref{P''}) this additional term is of the same order ${\cal O}(1)$
as those which we neglected. Unfortunately we know from the analysis of the 
real eigenvalue case as reviewed  
in \cite{KFrev} that terms of this order are 
important and do contribute. Our method is therefore not easily applicable to 
the chiral case.

%%%%%%%%%%%%%%%%%%%%%%%%%%%%%%%%%%%%%%%%%%%%%%%%%%%%%%%%%%%%%%%%%%%%
\sect{Conclusions}\label{con}   

We have shown that universality in random matrix models holds also for 
correlation functions of eigenvalues in the complex plane. 
The proof we have presented holds in the limit of weakly non-Hermitian matrices
where the parameter $\al$ measuring the non-Hermiticity times the size of the 
matrices $N$ is kept fixed when $N$ goes to infinity. Furthermore, we have 
restricted ourselves to the microscopic limit at the origin where complex 
eigenvalues at zero are magnified. This limit is particularly important in the 
application to QCD as the spectral density (of Dirac eigenvalues) directly 
measures the chiral condensate, the order parameter of chiral symmetry 
breaking. 

We have shown universality by analyzing the asymptotics of orthogonal 
polynomials in the complex plane using factorization in the large-$N$ limit. 
This lead us to a universal differential 
equation exactly as in the case of real eigenvalues. The universality of the 
kernel of polynomials and thus of all correlation functions then followed. 
The inclusion of an arbitrary number of Dirac mass terms was done by expressing
massive through massless correlators.

Our results present only a first step towards the  universality of all complex 
eigenvalues correlations. This is not only due to the fact that the number of 
symmetry classes is by far larger than that of matrix models with 
real eigenvalues. The exist also 
two fundamentally different large-$N$ limit for 
matrices with complex eigenvalues: the limit of weak and strong non-Hermiticity
where we have only investigated the former. The limit of weak non-Hermiticity 
is particularly important as it smoothly connects the know classes of 
real and strongly non-Hermitian eigenvalue correlations, by taking the limits 
$\al\to 0$ and $\al\to\infty$ respectively. These two classes 
would otherwise be completely disjoint. 
In fact by taking the limit $\al\to\infty$ of our universal, weakly 
non-Hermitian correlators we have a good argument in favor of universality 
at strong non-Hermiticity as well. 

Apart from the two classes of non-Hermiticity different correlations may also 
be found in other regions of the spectrum, in the bulk away from the 
origin and at the edge.  
Another question is that of macroscopic universality, which holds for all 
connected two- and higher $k$-point correlation functions of real eigenvalues. 
Such finding may have important consequences in phenomena such as 
wetting in two dimensions related to the inverse potential 
problem for analytic curves in the complex plane.

However, we consider universality of chiral complex matrix models, as very 
recently introduced by the author, as most urgent. Such models are aimed to 
describe the microscopic fluctuations of Dirac operator eigenvalues in 
four dimensional QCD which are complex due to the presence of a chemical 
potential. To answer that question we need to know subleading corrections in 
$N$ to the differential equation of 
polynomials. We leave this task for future plans.

\indent

\noindent
\underline{Acknowledgments}: 
I am indebted to P. Di Francesco, E. Kanzieper 
and G. Vernizzi for enlightening discussions and correspondence.
This work was supported by the European network on ``Discrete Random 
Geometries'' HPRN-CT-1999-00161 (EUROGRID).

%%%%%%%%%%%%%%%%%%%%%%%%%%%%%%%%%%%%%%%%%%%%%%%%%%%%%%%%%%%%%%%%%%%%%%%%%%%%%

\end{document}